\documentclass[conference]{IEEEtran}
\IEEEoverridecommandlockouts
\ifCLASSOPTIONcompsoc
\usepackage[caption=false,font=normalsize,labelfon
t=sf,textfont=sf]{subfig}
\else
\usepackage[caption=false,font=footnotesize]{subfi
g}
\fi
\usepackage{cite}
\usepackage{amsmath,amssymb,amsfonts}
\usepackage{algorithmic}
\usepackage{graphicx}
\usepackage{xcolor}
\usepackage{textcomp}
\usepackage{xcolor}
\usepackage{tabularx}
\usepackage{multirow}
\usepackage{float}
\usepackage{caption}
\usepackage{lipsum}
\usepackage{physics}
\def\BibTeX{{\rm B\kern-.05em{\sc i\kern-.025em b}\kern-.08em
    T\kern-.1667em\lower.7ex\hbox{E}\kern-.125emX}}
\begin{document}

\title{A Universal Quantum Technology Education Program \\
}
\makeatletter
\newcommand{\linebreakand}{%
  \end{@IEEEauthorhalign}
  \hfill\mbox{}\par
  \mbox{}\hfill\begin{@IEEEauthorhalign}
}
\makeatother


\makeatletter
\def\endthebibliography{%
  \def\@noitemerr{\@latex@warning{Empty `thebibliography' environment}}%
  \endlist
}
\makeatother

\author{\IEEEauthorblockN{Sanjay Vishwakarma}
\IEEEauthorblockA{\textit{Independent Researcher} \\
California, USA \\
svishwak@alumni.cmu.edu}
\and
\IEEEauthorblockN{Shalini D}
\IEEEauthorblockA{\textit{Universidad Internacional Menendez Pelayo} \\
Madrid, Spain \\
shalinibabu98@gmail.com}
\and
\IEEEauthorblockN{Srinjoy Ganguly}
\IEEEauthorblockA{\textit{Woxsen University} \\
Hyderabad, India \\
srinjoy.ganguly@woxsen.edu.in}
\linebreakand
\IEEEauthorblockN{Sai Nandan Morapakula}
\IEEEauthorblockA{\textit{Karunya Institute of Technology and Science} \\
Coimbatore, India\\
sainandanm2002@gmail.com}
}

\maketitle

\begin{abstract}
Quantum technology is an emerging cutting-edge field which offers a new paradigm for computation and research in the field of physics, mathematics and other scientific disciplines. This technology is of strategic importance to governments globally and heavy investments and budgets are being sanctioned to gain competitive advantage in terms of military, space and education. Due to this, it is important to understand the educational and research needs required to implement this technology at a large scale. Here, we propose a novel universal quantum technology master's curriculum which comprises a balance between quantum hardware and software skills to enhance the employability of professionals thereby reducing the skill shortage faced by the academic institutions and organizations today. The proposed curriculum holds the potential to revolutionize the quantum education ecosystem by reducing the pressure of hiring PhDs faced by startups and promoting the growth of a balanced scientific mindset in quantum research.
\end{abstract}


\section{Introduction}
The development of quantum computing started towards the early 20th century. With Max Planck's hypothesis in the 1900's era of quantum theory had begun. The contributions of Albert Einstein's photoelectric effect in 1905, De Broglie in 1924 proving the particle nature of light, Heisenberg's uncertainty principle, Erwin Schrodinger's formulation of quantum mechanics based on waves, Max Born's interpretation of wave function laid the foundation for understanding the quantum world. The main breakthrough happened with the development of quantum algorithms. Deutsch algorithm has theoretically proved more efficient than a classical algorithm, Peter Shor's 1994 factoring algorithm \cite{Shor_1997} will make a revolution in cryptography when implemented in a large-scale quantum computer, and Grover algorithm's \cite{grover1996fast}  ability to  search in the unsorted database in \(\sqrt{N}\) tries is considered to be faster than classical counterpart, VQE's contribution in quantum chemistry \cite{Peruzzo_2014} are astonishing development in the quantum field. 

Apart from the above-mentioned points, a few other reasons to move from classical computing are three quantum mechanical properties. These are Superposition, Interference and Entanglement. Superposition creates \(2^n\) dimensions where an increase in qubits has an exponential increase in dimensions, Interference leverages parallel computation, and Entangled qubits properties are known if one of the qubit's properties is known irrespective of how far the qubits are placed. Along with it speed, efficiency, solving complex problems, data security, scientific and medical research, and advancement in AI are also the reasons to shift to quantum computing. There are many opportunities to start as a complete beginner to get a glimpse. The Coding School is sponsored by IBM which is a complete 8-month dedicated course to give a solid understanding of The Introduction to Quantum Computing, QWorld, QOSF, and IBM Spring and Fall Challenge.IBM certified Qiskit developer exam and IBM Qiskit advocate are the world's only official exams organised by IBM to test the knowledge of basic to medium computing skills.  

Even though the mentioned opportunities are a good start, the depth of subject knowledge will be developed only by enrolling in formal education. A dedicated syllabus with an enrichment of varied research areas must be curated to create a strong foundation for the research mindset. Along with it, students need to be equipped enough to read research papers in a strategic way to get the most out of them. Once they are able to identify the areas of development quantum workforce will be extended to most of the fields. There might be a day in the future when we will use tech gadgets in our daily routine \cite{pallotta2022bringing}. As schools play a major role in one's education, we strongly hope that a lot of schools start inculcating quantum subjects such as quantum physics, and basic quantum chemistry into their high school curriculum \cite{MullerMishina_2021}. Thus, quantum computing is a promising field with rich resources and a developing research mindset. Once set, quantum will outsource present technologies and the world will reach its peak scientific advancement \cite{2021}. 

In this paper, we propose a novel universal quantum education curriculum for master's level degree programs. We start defining the course framework in section II, followed by a detailed description of the subjects and the semester-wise categorification of the subjects in section III. In section IV, we throw light on the industrial aspect and global outlook of our curriculum and its need.

\section{Course Framework}
The two-year master's program is designed to equip all students with the essential knowledge required to grow and sustain in the field of quantum science and technology. In this section, we introduce the course framework of this program. The pedagogy is built on a number of fundamental ideas, such as the following:
\begin{itemize}
  \item A variety of diversified learning opportunities that support the development of deep conceptual understanding and improve students' capacity for successful application of technical and professional skills.
  \item The course is designed in such a way that all the students are taught most of the modules using problem-based learning \cite{tang2021teaching}.
  \item As quantum computing is a diverse field that has wide range of applications, the course offers specializations to students based on their interests and future goals.
\end{itemize}

The Universal Quantum Technology Education Program's (UQTEP) curriculum is created to be at the cutting edge of current quantum technology research trends. The curriculum is designed to guarantee students are not only well-versed in established quantum concepts but are also conversant with the most recent advances in quantum science and technology since it recognizes the rapidly dynamic nature of the field. A detailed list of master's programs in quantum science and technology and their overview is discussed in the paper \cite{Aiello_2021}. \\
Modern research issues are incorporated into the course modules as one of the main approaches to promoting this relationship. For instance, students are introduced to quantum error correction and fault tolerance in the quantum information and computing module; these topics are crucial in the field's ongoing research. In line with the increased interest in these fields in both academia and industry, the quantum software development and simulation module also contains instructions on quantum algorithms and quantum machine learning.

The UQTEP also emphasizes the importance of applying quantum technologies in real-world settings. Students are encouraged to apply the theoretical information \cite{Singh_2007} they have acquired to actual quantum problems through practical laboratory sessions and a sizable research project in the second year. This offers students a priceless chance to interact with current research questions and acquire hands-on experience in the industry.\\
Additionally advantageous to the curriculum are the close ties to the quantum research community. Many of the program's teachers are now engaged in quantum technology research, ensuring that the teaching is guided by the most recent advancements in this field. Regular guest lectures from industry experts offer further perspectives on the current condition of the industry and the practical uses of quantum technology.

Additionally, the program aggressively promotes student participation in the larger quantum research community. Students can gain first-hand knowledge of the quantum research and development process through opportunities for internships or placements in research institutions or quantum technology enterprises. The relationship between students' academic work and the larger research community is further strengthened by encouraging them to attend and present at conferences on quantum technologies. In conclusion, there is a strong connection between the UQTEP curriculum and contemporary quantum technology research trends. The program makes sure that its graduates are prepared to contribute to this interesting and quickly developing topic by including the most recent research advancements in the curriculum and offering chances for practical application and involvement with the quantum research community.

The Universal Quantum Technology Education Programme (UQTEP) acknowledges that quantum technology transcends conventional divisions between physics, computer science, engineering, and mathematics and is intrinsically interdisciplinary \cite{MullerMishina_2021}. This is why the program puts a lot of focus on interdisciplinary learning and cooperation, giving students lots of chances to interact with others and ideas from many sectors. The program's core curriculum, which covers a variety of subjects from quantum physics \cite{Krijtenburg_Lewerissa_2017} to quantum computing and software development, reflects the interdisciplinary nature of the UQTEP. Students obtain a wide understanding of quantum technology that crosses several traditional subjects by studying these various fields. Additionally, the programme encourages students to choose elective courses that are unrelated to their primary field of study. These elective courses give students the chance to explore subjects that may not be directly relevant to their chosen career path, but are nonetheless crucial for gaining a comprehensive grasp of quantum technology. To acquire exposure to the software side of quantum technology, a student specializing in the hardware pathway can decide to take a module in quantum machine learning or quantum cryptography.

The UQTEP encourages interdisciplinary cooperation through a number of initiatives outside of the traditional curriculum \cite{economou2022hello}. For instance, group projects are a crucial component of the curriculum. These projects frequently involve interdisciplinary teams of students solving challenging issues, promoting cooperation, and allowing students to benefit from one another's experience. These projects foster a rich interdisciplinary learning environment by allowing students from the software track to collaborate with peers from the hardware and superconducting circuits pathways. Additionally, the second-year research project of the programme provides a fantastic opportunity for multidisciplinary study. Each student's project is in keeping with their chosen career path, however due to the nature of quantum technology research, cross-disciplinary collaboration is frequently necessary. For instance, a student working on a project with a hardware focus might team up with academics who are experts in quantum materials or algorithms. Last but not least, the UQTEP strongly promotes student interaction with the larger academic and business world, both inside and outside the realm of quantum technology. This could be participating in multidisciplinary seminars and workshops, doing an internship with a variety of businesses, or working on projects or conducting research with other researchers or business experts. In conclusion, the Universal Quantum Technology Education Program goes beyond traditional disciplinary boundaries, fostering an interdisciplinary learning environment that mirrors the diverse and interconnected nature of quantum technology research and development. Through a combination of a broad curriculum, collaborative projects, and engagement with the wider academic and industry community, the program prepares students to thrive in the interdisciplinary world of quantum technology.\\

\section{Compulsory Modules and Electives}

In order to create a world-class industry and academia standard curriculum, it is necessary to define a few subjects which are of relevance to the program and provides a unique way to bind the topics together. We define several subjects to be taught in our proposed master's program which defines hardware and software aspects of quantum technology. The curriculum is designed taking the reference of the research papers \cite{perron2021quantum}, \cite{Kaur_2022}, \cite{Bonacci_2020}, \cite{zuccarini2023promoting}, \cite{doi:10.1080/09500690110073982} \cite{Asfaw_2022}. The paper \cite{economou2020teaching}, provided a customizable outreach program that is accessible to students in high school and the early stages of their undergraduate studies, requires no sophisticated mathematics, and can be utilized to introduce concepts to students in a short amount of time. \cite{Meyer_2022} reports on the results of a survey of instructors instructing introduction to courses at universities across the US, mostly at the undergraduate or hybrid undergraduate/graduate level, and by conducting follow-up focus interviews with six specific teachers. Teaching with the help of visualization techniques is also a great advantage as it helps students understand and grasp the concepts of quantum computing better \cite{Nita_2021}. We define details of some of the important subjects proposed in our curriculum and explain their significance.

\subsection{Introduction to Quantum Algorithms}
The subject "Introduction to quantum algorithms" provides a foundational grasp of the algorithms employed in the field of quantum computing. Quantum algorithms are specifically intended to take advantage of quantum systems' unique features in order to solve computing problems more efficiently than classical algorithms. Students will study the main concepts and principles behind quantum algorithms in this class. They will investigate several algorithms that use quantum phenomena including superposition, entanglement, and interference to execute calculations. These algorithms are very different from classical algorithms, which are based on classical binary logic and work with classical bits. The basic building blocks of quantum algorithms, such as quantum gates, qubits (quantum bits), and quantum circuits, are often included in the curriculum. Students will learn about fundamental algorithms such as Shor's algorithm for factoring huge numbers and Grover's algorithm for unstructured search. They will comprehend the mathematics and ideas underpinning these algorithms, as well as their potential applications in cryptography, optimization, simulation, and machine learning.\\
Furthermore, students will investigate quantum algorithm design concepts, learning how to identify issues that can benefit from quantum computation and build appropriate solutions. They will learn about quantum oracle creation, quantum Fourier transforms, amplitude amplification and other fundamental quantum algorithm approaches. Throughout the course, students may be introduced to quantum computing programming languages and frameworks such as Qiskit, Microsoft Quantum Development Kit (Q\#), and others. Hands-on exercises and projects that allow students to develop and simulate quantum algorithms may be incorporated to reinforce theoretical notions. Students should have a good foundation in quantum algorithms at the end of the course, allowing them to understand the unique capabilities and promise of quantum computing. They will be prepared to analyze problems, identify appropriate quantum algorithms, and contribute to quantum technological advancements.
\subsection{Semiconductor Devices}
The subject "Semiconductor Devices" is concerned with the underlying principles, operation, and applications of semiconductor devices in the realm of electronics. It gives pupils a thorough understanding of the building components that constitute the foundation of modern electrical systems. Students will learn about the physics and engineering of semiconductor devices, which are primarily based on the properties of semiconducting materials like silicon. They will study about the behaviour of electrons and holes in semiconductors, pn junction creation, and carrier transport concepts in these devices.\\
The curriculum often includes diodes, bipolar junction transistors (BJTs), metal-oxide-semiconductor field-effect transistors (MOSFETs), and other sophisticated devices such as photodiodes and optoelectronic devices. Students will investigate the principles of operation, device properties, and applications of these devices in various electronic systems. Key subjects in the discipline may include diode and transistor biassing and small-signal analysis, the study of load lines and operating zones, amplifier architectures, frequency response, and basic device fabrication procedures. Students will also learn about the constraints and trade-offs that go into the design and operation of these devices.\\
Furthermore, advanced semiconductor devices such as integrated circuits (ICs) and sensors may be covered. Students will learn about IC fabrication techniques, device scaling, and integrating several components on a single chip. They might also learn about design considerations for sensors that use semiconductor materials to detect physical quantities. Students may participate in practical exercises, laboratory work, and projects throughout the course to reinforce theoretical concepts and acquire hands-on experience with semiconductor devices. They may also investigate new trends and advances in semiconductor technology, such as nanoscale devices, power electronics, and semiconductor materials other than silicon.\\
Students should have a comprehensive understanding of semiconductor devices, their underlying concepts, and their significance in electronic systems at the end of the course. They will be able to analyse, design, and troubleshoot electronic circuits that make use of these devices. This foundation will be helpful for future studies or professions in electrical engineering, the semiconductor industry, integrated circuit design, and electronic system development.
\subsection{Laboratory and Research Skills}
The information needed to work on cutting-edge quantum technology research and development is covered in "Laboratory and Research Skills for Quantum Technology". The course offers practical lab instruction where students operate, manipulate, and conduct experiments with quantum systems. This could entail performing experiments in quantum physics, establishing quantum communication networks, or working with quantum computing technology.

Equally important are research abilities. These include the capacity to plan experiments, examine data, create and test theories, and stay abreast of the quantum technology field's quick evolution. Students are better equipped to progress in quantum technology with these skills, whether they choose to work in academic research or the private sector.

This topic has a dramatic impact on a global education programme on quantum technologies. It enables students to translate abstract quantum principles into concrete technological advances by bridging the gap between theoretical understanding and actual implementation. The development of quantum technology and the training of the following generation of quantum scientists and engineers depend on this fusion of theory and practice.

\subsection{Introduction to Quantum Photonics}

The crucial topic "Introduction to Quantum Photonics" explores the relationship between quantum mechanics and light, or more particularly, photons. The focus of the study is on the fundamental ideas underlying the production, control, and measurement of photons within the framework of quantum mechanics. As they pertain to photonics, this includes comprehending phenomena like quantum entanglement, quantum superposition, and quantum interference.

The practical use of these ideas in quantum sensing, quantum computing, and quantum communication will also be demonstrated to the students. For instance, scientists might study quantum teleportation, quantum cryptography, photonic quantum bits (qubits), and the creation of quantum networks.

As photonics is a fast-evolving topic with many potential applications in quantum technology, it plays a vital part in a programme for universal quantum technology education. It serves as a foundation for comprehending the processing and transmission of quantum information, making it a prerequisite for anyone wishing to specialise in quantum technology. Furthermore, a solid understanding of quantum photonics can lead to a variety of research and job prospects because optical systems are frequently employed to create quantum technologies.

\begin{center}	
\begin{tabular}{| >{\centering\arraybackslash}m{5.5em} | >{\centering\arraybackslash}m{3cm}| m{3cm} |}
	\hline
	\textbf{Master Year} & \textbf{Period} & \textbf{Subjects} \\
	\hline
	\multirow{2}{*}{1st} & Semester 1 & \textbf{Core:} Quantum Mechanics, Quantum Information and computing, Intro to quantum algorithms, Quantum optics and quantum matter. \textbf{Electives:} statistical mechanics, Quantum electrodynamics, Introduction to data science, Group theory, Solid state system for Quantum information  \\
	\cline{2-3}
	 & Semester 2 & \textbf{Core:} Advanced quantum mechanics, Laboratory and research skills, Semiconductor devices, Nanotechnology, Programming for physics. \textbf{Electives:} Fundamentals of solid state materials, Lab in nanoelectronics, Molecular electro dynamics, Semiconductor for physics and light-matter interaction, Introduction to quantum photonics.

 \\
	\hline
	\multirow{2}{*}{2nd} & Semester 3 &\textbf{Hardware:} Photonics, Semiconductor. \textbf{Software:} Programming and qiskit, Pennylane

  \\
	\cline{2-3}
	& Semester 4 & \textbf{Hardware and software:} Research project, Thesis or internship 
.\\
	\hline
\end{tabular}
\end{center}

\section{International Industry Focused Curriculum}

In the current landscape, we can observe a noticeable talent shortage in the quantum technology market primarily due to an absence of a universal quantum education curriculum which combines industry-relevant skills with an innovation-driven mindset. Due to this, most industries are focusing on hiring PhDs who are already trained to innovate and skilled in the execution of multiple complex research projects relevant to the quantum industry. But the hiring of such candidates comes at a cost of financial strain on the startups that are facing budget delays and tightening. 

In order to resolve the challenges, a world-class and industry-standard curriculum is required at the level of a master's degree because such a program is closer to PhDs and brings more specialisation of topics opening pathways to different career streams. This approach proves to be more economical and time-saving since it does not necessitate long study time and research. Furthermore, these master's specialisations provide in-depth knowledge about a particular technology that is in line with the job descriptions given by organisations today. Additionally, the industry-oriented curriculum ensures a strategic move towards achieving a smoother transition into the quantum workforce, irrespective of geographical boundaries, thus optimizing talent utilisation in this niche field,

The cutting-edge master's degree programs must maintain a balance between the theoretical and practical aspects of the subjects by continuously following international and industry-relevant standards. The curriculum should be comprehensive enough to cover the skills and criteria required by companies globally with ample internships and research projects. This can prepare the students for multiple roles in quantum industries such as research software engineers, nanofabrication engineers, research scientists in quantum chemistry or algorithms, measurement and design engineers and so on. This promotes a positive factor of employability and the shortening of skill-gap present among emerging professionals today.

While developing a cutting-edge curriculum, it is important to advocate for equal opportunities, inclusion, and diversity for the marginalized sections of society that can ensure a clear career path for the students and professionals in the quantum space. Placing a strong emphasis on professional development will enable organizations to foster a culture of growth and will aid in the cultivation of professional allegiance to that respective industry. 

A quantum-ready workforce has the potential to serve not only the commercial industries but also the government agendas for national security and intelligence. It will be important to upskill scientists and professionals who could contribute to the development of the nation's defence and enable technological breakthroughs. Therefore, there is an urgent call for industry-relevant quantum technology syllabi so as to achieve global competitiveness in this sector and enhance the education sector of this field.

\section{Conclusion}

The revolution of quantum technology is currently in its very early stages and it is difficult to predict the far future of this technology. However, by observing the current pace of research, it is possible that we will witness truly amazing technological breakthroughs in the upcoming years. In order to have a successful quantum ecosystem, the most significant role is that of having cutting-edge quantum education and research that enables professionals to be at the forefront of their careers and businesses. In this paper, we proposed novel ideas for the development of a universal master's degree program which will allow students and professionals to gain industry-relevant niche skills and provides them with an overall overview of the current research and business scenarios happening in the quantum space along with career growth. We hope that our proposed master's curriculum is adopted by academic institutions globally to make their system robust and competitive. Furthermore, we are ambitious that our universal methodology of syllabus is able to solve the pressing needs of the current quantum education system, and empower and inspire students and professionals to pursue their careers in this futuristic field.

\section*{Acknowledgment}

S.G. acknowledges the support received from Woxsen University to develop the curriculum for quantum education. S.N.M. is grateful to Karunya Institute of Technology and Science for providing the opportunity to participate in the syllabus design process. 

\bibliographystyle{IEEEtran}
\bibliography{References}

\end{document}